\documentclass[twocolumn,trackchanges]{aastex7}

\usepackage{amsmath}
\usepackage{graphicx}
\usepackage{subfigure}
\usepackage{enumerate}
\usepackage[figuresright]{rotating}
\usepackage{longtable}
\usepackage{threeparttablex}
\usepackage{tablefootnote}
\usepackage{multirow}
\usepackage{enumitem}
\usepackage[normalem]{ulem}
\newcommand{\ESN}{E_{\rm SN}}

\newcommand{\ping}[1]{{\color{black} #1}} 
\definecolor{forestgreen}{rgb}{0.13, 0.55, 0.13}
\newcommand{\zy}[1]{\textcolor{black}{#1}}
\newcommand{\comment}[1]{{\color{black} #1}}
\newcommand{\commentv}[1]{{\color{black} #1}}
\newcommand{\commentvv}[1]{{\color{black} #1}}
\begin{document}

\title{Absence of Radio Emission Reveals an Exceptionally Weak Explosion of the putative Historical Supernova Pa 30}

\author[orcid=0000-0001-5684-0103,sname='Shao']{Yi-xuan Shao}
\affiliation{School of Astronomy and Space Science, Nanjing University, Nanjing, 210023, People's Republic of China}
\affiliation{Key Laboratory of Modern Astronomy and Astrophysics, Nanjing University, Ministry of Education, Nanjing, 210023, People's Republic of China}
\email[hide]{yixuan@smail.nju.edu.cn} 

\author[orcid=0000-0002-5683-822X,sname='Zhou']{Ping Zhou}
\affiliation{School of Astronomy and Space Science, Nanjing University, Nanjing, 210023, People's Republic of China}
\affiliation{Key Laboratory of Modern Astronomy and Astrophysics, Nanjing University, Ministry of Education, Nanjing, 210023, People's Republic of China}
\email[show]{pingzhou@nju.edu.cn}

\author[orcid=0000-0002-9392-547X,sname='Zhang']{Xiao Zhang}
\affiliation{School of Physics \& Technology, Nanjing Normal University, No.1 Wenyuan Road, Nanjing 210023, China}
\email[show]{xiaozhang@njnu.edu.cn}

\author[orcid=0000-0002-7299-2876,sname='Zhang']{Zhi-Yu Zhang}
\affiliation{School of Astronomy and Space Science, Nanjing University, Nanjing, 210023, People's Republic of China}
\affiliation{Key Laboratory of Modern Astronomy and Astrophysics, Nanjing University, Ministry of Education, Nanjing, 210023, People's Republic of China}
\email[hide]{zzhang@nju.edu.cn}

\author[orcid=0000-0002-4753-2798,sname='Chen']{Yang Chen}
\affiliation{School of Astronomy and Space Science, Nanjing University, Nanjing, 210023, People's Republic of China}
\affiliation{Key Laboratory of Modern Astronomy and Astrophysics, Nanjing University, Ministry of Education, Nanjing, 210023, People's Republic of China}
\email[hide]{ygchen@nju.edu.cn}

\author[orcid=0000-0002-2342-9956,sname='Han']{Qin Han}
\affiliation{Mullard Space Science Laboratory, University College London, Holmbury St. Mary, Surrey RH5 6NT, UK}
\email[hide]{qin.han.21@ucl.ac.uk}

\author[orcid=0000-0003-3010-7661,sname='Li']{Di Li}
\affiliation{New Cornerstone Science Laboratory, Department of Astronomy, Tsinghua University, Beijing 100084, China}
\affiliation{National Astronomical Observatories, Chinese Academy of Sciences, Beijing 100101, China}
\email[hide]{dili@tsinghua.edu.cn}

\author[orcid=0000-0002-0584-8145,sname='Li']{Xiang-Dong Li}
\affiliation{School of Astronomy and Space Science, Nanjing University, Nanjing, 210023, People's Republic of China}
\affiliation{Key Laboratory of Modern Astronomy and Astrophysics, Nanjing University, Ministry of Education, Nanjing, 210023, People's Republic of China}
\email[hide]{lixd@nju.edu.cn}

\author[orcid=0000-0002-3576-441X,sname='Weng']{Jian-Bin Weng}
\affiliation{School of Astronomy and Space Science, Nanjing University, Nanjing, 210023, People's Republic of China}
\affiliation{Key Laboratory of Modern Astronomy and Astrophysics, Nanjing University, Ministry of Education, Nanjing, 210023, People's Republic of China}
\email[hide]{waynewengjianbin@outlook.com}

\author[orcid=0000-0003-2506-6906,sname='Shao']{Yong Shao}
\affiliation{School of Astronomy and Space Science, Nanjing University, Nanjing, 210023, People's Republic of China}
\affiliation{Key Laboratory of Modern Astronomy and Astrophysics, Nanjing University, Ministry of Education, Nanjing, 210023, People's Republic of China}
\email[hide]{shaoyong@nju.edu.cn}

\begin{abstract}
We present the first deep radio continuum observations of Pa 30, a nebula hosting a unique optical source driven by an ultrafast outflow \commentv{with} a velocity of 16,000 km~s$^{-1}$. The nebula was proposed to be the remnant of a white dwarf merger that occurred in 1181CE.  We report no detection of the radio diffuse emission from Pa~30 or radio emission from the central source, setting $3\sigma$ upper limit \commentv{flux densities} of \commentv{$0.84\,\rm mJy$ and $0.29\,\rm mJy$} at 1.5~GHz and 6~GHz, respectively, for Pa~30. The radio surface brightness of Pa~30 is $\sim 3$ orders of magnitude smaller than that of typical supernova remnants (SNRs) with comparable angular size. If Pa 30 is an SNR, our observations show it to be the faintest known in the radio band. Considering that 10\% \commentv{of the supernova (SN) kinetic} energy is transferred to cosmic rays (CRs), the absence of radio synchrotron emission suggests that the \commentv{SN kinetic} energy \commentv{$\lesssim3\times 10^{47}(B/10~\mu\rm G)^{-1.65}$~erg}, which is 3 to 4 orders of magnitude lower than that of typical SNRs and the lowest measured among Galactic SNRs. There is also an indication of inefficient CR acceleration for this source. The low \commentv{SN kinetic energy} either implies the potential existence of many more radio-faint, sub-energetic SNRs in our Galaxy or challenges the SNR interpretation of Pa~30.

\end{abstract}

\section{Introduction}

\ping{Recent observations have revealed that the properties of supernova remnants (SNRs) are remarkably different from uniform. While the canonical \commentv{SN kinetic} energy $\ESN$ is believed to be around $10^{51}$~erg, some SNRs are found to have much lower\commentv{-}energy SN explosions, such as the Crab Nebula with $\ESN \lesssim 10^{50}$~erg \citep[e.g.,][]{smith13,yang15}, DA 530 with $\ESN \lesssim 10^{50}$~erg \citep[e.g.,][]{Jiang+2007ApJ_DA530} and RCW~103 with $\ESN\lesssim 0.3$--$1\times 10^{50}$~erg \citep{braun19,zhou19}. The broad \commentv{kinetic energy} range found in SNRs \citep[0.03--$6\times 10^{51}$~erg][]{leahy20} suggests that there are explosion mechanisms beyond the benchmark Type Ia and typical core-collapse supernova explosion models. Finding and investigating peculiar SNRs helps to understand the various SN explosion mechanisms and refine the impact of SNe feedback on galaxies. 

\begin{figure*}[htp]
\centering
\includegraphics[width=1\textwidth]{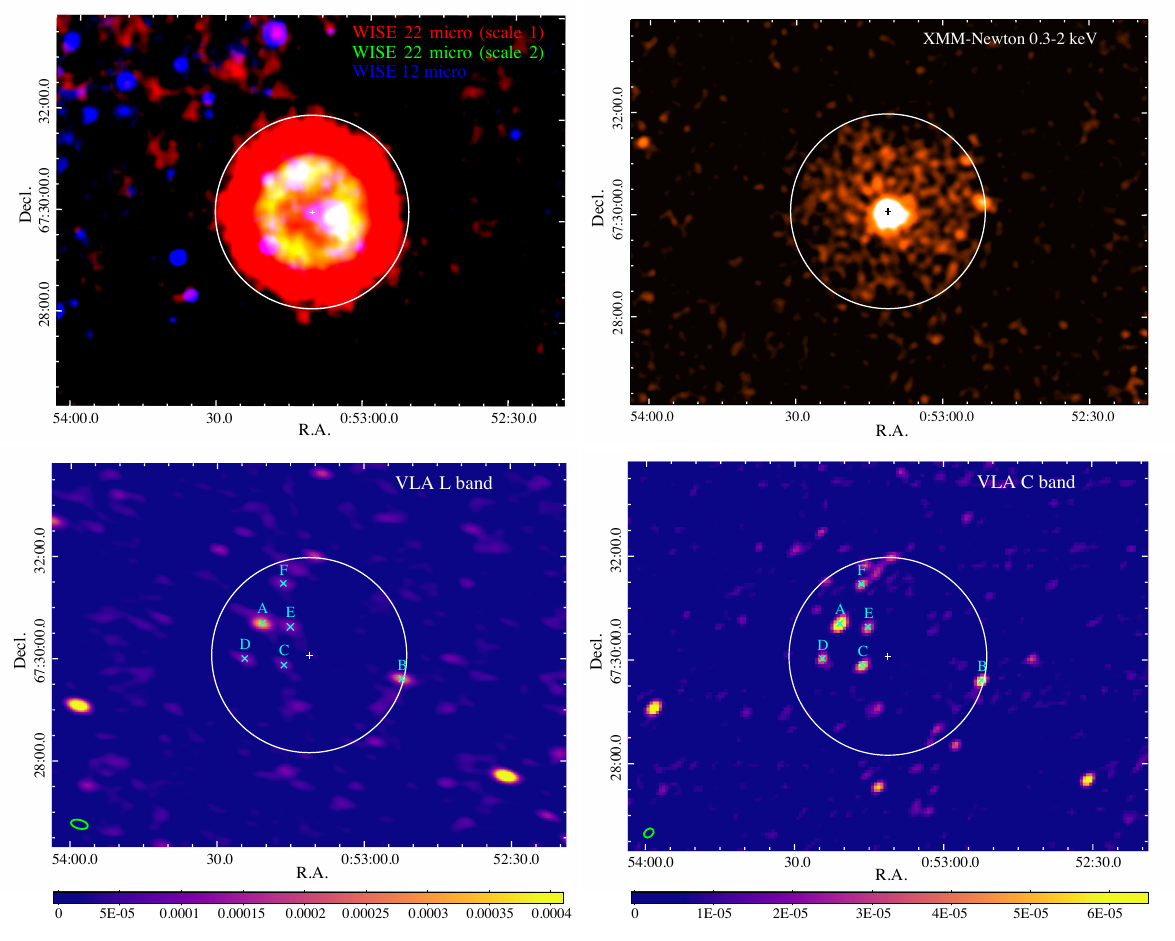}
\caption{\textit{Upper left:} Wide-field Infrared Survey Explorer (WISE) image of the Pa~30 field, red and green show $22\ \mu m$ under two intensity scales, and blue shows $12\ \mu m$. \textit{Upper right:} XMM-Newton X-ray flux image in 0.3--2 keV. \textit{Lower left:} VLA $1.5\, \rm{GHz}$ (L band) image. \textit{Lower right:} VLA $6.0\, \rm{GHz}$ (C band) image. The white circles are $114.5''$ in radius centered on the central star J005311 \citep[$\rm R.A.=00^{h}53^{m}11.2^{s}, Decl.=+67^{\circ}30'2.4''$,][]{Fesen+2023ApJ...945L...4F}. \commentv{The cyan `x' markers represent the location of the compact sources.} 
The green ellipse in the lower-left corner of both VLA images represents the synthesized beam. The colorbars show the flux density in units of $\rm Jy/beam$.}
\label{Fig:4in1}
\end{figure*}

\begin{table*}[htp]
    \centering
    \caption{\textbf{Observational information and flux density}}
    \begin{tabular}{ccccccc}
    \hline
    \hline
    \multirow{2}{*}{Region} &R.A.&Decl.& Frequency & Flux Density & \multirow{2}{*}{Frequency Spectral Index}\\
    &(J2000)&(J2000)&(GHz)&($\rm mJy$)&\\
    \hline
    \multirow{2}{*}{Pa~30}&\multirow{2}{*}{$\rm00^{h}53^{m}11.2^{s}$}&\multirow{2}{*}{$+67^{\circ}30'2.4''$}& 1.5 &$<0.84$\tablenotemark{*}&\multirow{2}{*}{-}\\
    &&&6.0&$<0.29$\tablenotemark{*}&\\
    \hline
    \multirow{2}{*}{J005311}&\multirow{2}{*}{$\rm00^{h}53^{m}11.2^{s}$}&\multirow{2}{*}{$+67^{\circ}30'2.4''$}& 1.5 &$<0.06$\tablenotemark{*}&\multirow{2}{*}{-}\\
    &&&6.0&$<0.01$\tablenotemark{*}&\\
    \hline
    \multirow{2}{*}{A}&\multirow{2}{*}{$\rm00^{h}53^{m}20.8^{s}$}&\multirow{2}{*}{$+67^{\circ}30'42.5''$}& 1.5 &$0.53\pm0.14$&\multirow{2}{*}{$-1.11\pm0.19$}\\
    &&&6.0&$0.11\pm0.01$&\\
    \hline  
    \multirow{2}{*}{B}&\multirow{2}{*}{$\rm00^{h}52^{m}52.3^{s}$}&\multirow{2}{*}{$+67^{\circ}29'36.9''$}& 1.5 &$0.34\pm0.06$&\multirow{2}{*}{$-1.00\pm0.17$}\\
    &&&6.0&$0.09\pm0.01$&\\
    \hline
    \multirow{2}{*}{C}&\multirow{2}{*}{$\rm00^{h}53^{m}16.4^{s}$}&\multirow{2}{*}{$+67^{\circ}29'53.1''$}& 1.5 &$0.11\pm0.04$&\multirow{2}{*}{$-0.23\pm0.28$}\\
    &&&6.0&$0.08\pm0.01$&\\
    \hline
    \multirow{2}{*}{D}&\multirow{2}{*}{$\rm00^{h}53^{m}24.4^{s}$}&\multirow{2}{*}{$+67^{\circ}30'2.2''$}& 1.5 &$0.15\pm0.03$&\multirow{2}{*}{$-0.60\pm0.18$}\\
    &&&6.0&$0.06\pm0.01$&\\
    \hline
    \multirow{2}{*}{E}&\multirow{2}{*}{$\rm00^{h}53^{m}15.3^{s}$}&\multirow{2}{*}{$+67^{\circ}30'38.8''$}& 1.5 &$0.20\pm0.12$&\multirow{2}{*}{$-0.92\pm0.43$}\\
    &&&6.0&$0.06\pm0.01$&\\
    \hline
    \multirow{2}{*}{F}&\multirow{2}{*}{$\rm00^{h}53^{m}16.5^{s}$}&\multirow{2}{*}{$+67^{\circ}31'29.0''$}& 1.5 &$0.19\pm0.06$&\multirow{2}{*}{$-0.96\pm0.30$}\\
    &&&6.0&$0.05\pm0.01$&\\
    
    \hline
    \end{tabular}
    \tablenotetext{*}{$3\sigma$ upper limits.}
    \label{tab:VLA_result}
\end{table*}

Thermonuclear \commentv{SNe}, with Type Ia being the most well-known subgroup, are diverse, as demonstrated by both supernova observations \citep{Jha+2019NatAs_} and SNR studies \citep[see e.g.,][]{Zhou+2021ApJ...908...31Z,weng24}.}
White dwarfs (WDs) are created by stars with masses less than 8--10 $M_{\odot}$. When approaching or exceeding the Chandrasekhar mass limit, WDs in binary systems can lead to \commentv{thermonuclear} explosions as a type Ia SN disrupting the merger product, \comment{a type Ia SN with a neutron star (for massive WD binaries)}, or an accretion-induced collapse to form a neutron star \citep{Iben+1984ApJS...54..335I_typeIa,Pakmor+2011A&A...528A.117P_SNIa,king01,Saio+2004ApJ...615..444S,Shen+2012ApJ...748...35S}. However, there is increasing evidence that WDs may produce peculiar thermonuclear SNe with properties distinguished from normal Type Ia, such as the subluminous groups Type Iax, which may leave a stellar remnant behind \citep{Jha+2019NatAs_,Kromer+2013MNRAS.429.2287K} \ping{and Calcium-rich transients \citep{Perets+2010Natur_calcium,Kasliwal+2012ApJ_calcium}. 
SNR observations have identified two Galactic Type Iax SNR candidates, Pa~30 \citep{Oskinova+2020A&A} and Sgr A East near the Galactic center \citep{Zhou+2021ApJ...908...31Z}, and a Calcium-rich transient SNR candidate G306.3$-0.9$ \citep{weng22}.}

Pa~30, \ping{with a diameter of $\sim 220''$} \commentv{at a distance of $2.3\ \rm kpc$ \citep{Bailer-Jones2021AJ_distance},} is a newly discovered Galactic infrared (IR) nebula (a.k.a.\ IRAS 00500+6713) hosting a hot center object J005311 at R.A.=00h53m11.2s, Decl.=$+67^{\circ}30'2.1''$ \citep{Gvaramadze+2019Natur.569..684G}. 
\ping{The central hot source, with a temperature of \commentv{around $210000\, \rm K$}, drives ultra-fast winds of a velocity }$v_\infty\approx 16000\, \rm km\ s^{-1}$, suggesting a super-Chandrasekhar mass object due to a WD merger \citep{Gvaramadze+2019Natur.569..684G,Lykou+2023ApJ...944..120L}. 
\ping{The following up X-ray observations revealed O and C enhancements of the ejecta, supporting the merger of an ONe and a CO WD for creating Pa~30 \citep{Oskinova+2020A&A}.

The optical measurements show that Pa 30 is expanding with a peak velocity of $\approx 1400\, \rm km\ s^{-1}$ \citep{Ritter+2021ApJ...918L..33R,Fesen+2023ApJ...945L...4F,Cunningham2024ApJL_age} and imply an explosive event $\sim 10^3$~kyr ago. This young age, along with the nebula's position in the sky, establishes a connection between the nebula and the 1181 CE ``Guest star'' recorded in ancient Chinese and Japanese texts \citep{Ritter+2021ApJ...918L..33R,Schaefer+2023MNRAS}. This association is strengthened by recent Keck IFU observations, which provided a refined explosion time of $1152^{+77}_{-72}$~yr for Pa 30 \citep{Cunningham2024ApJL_age}.}
Moreover, Pa~30 reveals a remarkably peculiar morphology, containing an infrared (NIR) ring and radially aligned [S II] filaments pointing to the central object. This morphology is unlike any other \commentv{nova} or young Galactic \commentv{SNR} \citep{Ritter+2021ApJ...918L..33R,Fesen+2023ApJ...945L...4F}. 
\ping{Despite the young age, Pa~30 has not been detected in the radio band based on the shallow radio survey data (e.g., NVSS), setting the upper limit of radio flux density to $\sim4.1\ \rm mJy$ at the L band.}

\ping{Radio emissions in SNRs are synchrotron radiation generated by GeV-energy electrons, which can be easily accelerated in even old SNRs. Most SNRs in our Galaxy have been identified through radio surveys \citep{Green_catlog2025JApA...46...14G}, though a fraction are initially discovered through other wavebands. It is puzzling that Pa~30, which has an expansion velocity of $\sim 10^3~\rm km~s^{-1}$, is not bright in the radio band if it is indeed an SNR. Therefore, we conducted dedicated radio observations toward Pa~30, aiming to detect radio emission or study its puzzling properties in the radio band.}
In this \commentv{letter}, we report the first deep radio observation results of our VLA observations of Pa~30 in L and C bands. In Section \ref{sec:obs and data reduction} we describe our observations and the data reduction threads. The results of the new observation are presented in Section \ref{sec:results}. We discuss the emission models and their implications for different scenarios in Section \ref{sec:discussions}. Section \ref{sec:summary} is a summary of this work.

\section{Observations and Data Reduction}
\label{sec:obs and data reduction}

\zy{Our Pa~30 observations were conducted with the Very Large Array (VLA),
during August and October 2022 (22A-256; PI: Ping Zhou). The observations were
done with two epochs (executions), for $1.0$--$2.0$ GHz (L band) under C
configuration and $4.0$--$8.0\, \rm GHz$ (C band) under D configuration,
respectively. The choice for such configurations is optimised for their
comparable angular resolutions.  The largest angular scales are $\theta_{\rm
LAS}=970''$  and $\theta_{\rm LAS}=240''$ for L band and C band, respectively.
For both L- and C bands, the flux and bandpass calibrator was 3C~48, and the gain
calibrator was J0102+5824. We loop between Pa~30 and J0102+5824 every 15 and 10
minutes, for the L and C bands, with 90 seconds on the gain calibrator. The
on-source time was 85.5 minutes and 143.5 minutes at L band and C band,
respectively. }

\zy{ We processed the data using the standard VLA pipeline (ver 2024.1.0.8) with the
Common Astronomy Software Applications package \citep[CASA
v6.6.1;][]{CASA2022PASP..134k4501C}. After the standard pipeline processing, we applied additional manual flagging to remove low-level radio frequency
interference (RFI) and noisy scans on the source. 
}

\begin{figure*}[htp]
\centering
\includegraphics[width=\textwidth]{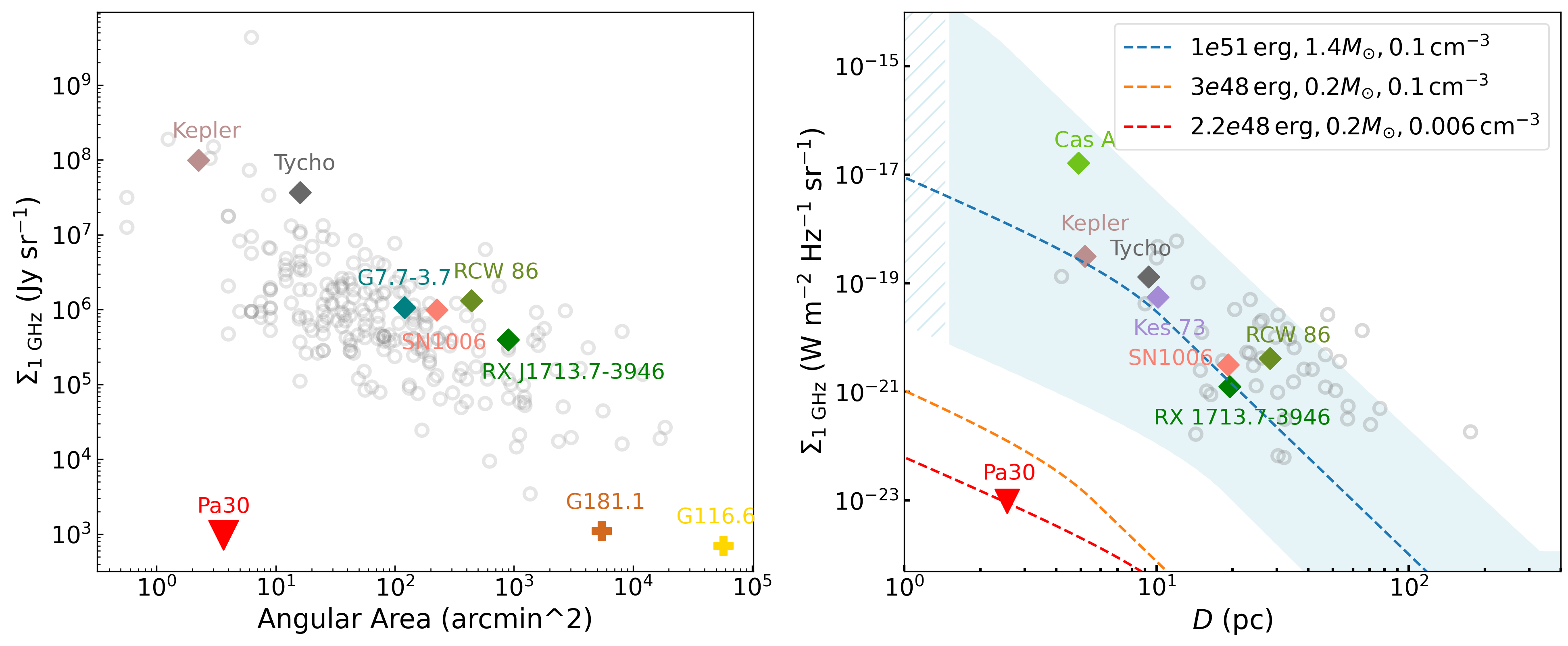}
\caption{\textit{Left:} The 1~GHz radio surface brightness vs.\ angular area diagram of \commentv{shell-type} SNRs, using the catalog by \cite{Green_catlog2025JApA...46...14G},\footnote{\url{https://www.mrao.cam.ac.uk/surveys/snrs/index.html}}. The diamond markers represent historical SNRs. The `+' markers show two newly found SNRs, G181.1+9.5 \cite{Kothes+2017A&A...597A.116K_G181.1} and G116.6-26.1 \cite{Churazov2022MNRAS.513L..83C_G116.6}) with low radio surface brightness. \textit{Right:} The $\Sigma$--D diagram, with the grey dots representing the \commentv{shell-type} SNRs with known distance taken from \cite{Pavlovic+2013ApJS.catalog}, and the diamond markers showing young SNRs with good distance measurements. The dashed lines show three exemplified cases with different \commentv{SN kinetic} energy ($E_{\text{SN}}$), ejecta mass ($M_{\rm ej}$), and ambient density ($n_0$), where $s=2.3$ \commentv{and $\epsilon_B=0.001$} is assumed. The blue shaded region shows the area predicted for most SNRs with \commentv{$s=2$--$2.5$}, $E_{\text{SN}}=10^{50}$--$2\times10^{51}\,\rm erg$, $M_{\rm ej}=1$--$4M_{\odot}$, $n_0=0.05$--$1\,\rm cm^{-3}$, and $\epsilon_{B}=0.001$--\commentv{$0.1$.}
The red triangle denotes the $3\sigma$ upper limit of Pa~30, extrapolated from 6 GHz
using a \commentv{frequency spectral index} of 0.65.}
\label{Fig:Surfbrightness_sigmaD}
\end{figure*}

\zy{We imaged the VLA data using the CASA task \textit{tclean}, with pixel sizes of
$1.8''$ and  $3''$ for L and C bands, respectively. The deconvolution adopts the standard gridding and a Briggs weighting with a robust parameter of 0.5, and multi-term multi-frequency synthesis \citep[MTMFS;][]{Rau+2011A&A...532A..71R} with two terms. The resulting synthesized beam sizes
are $20'' \times 11''$ (PA = $75^\circ$) and $13'' \times 9''$ (PA =
$-50^\circ$) in L band and C band, respectively. The fields of view (FOV) are
approximately $40.7'$ and $10'$ across, respectively. 
}

A few strong continuum sources outside the field of view produce ripples that
increase the final noise level, especially for the L-band data. Therefore, for
each of these sources, we shift the phase center to the target, clean, model
their visibility, and subtract it from the UV data. With the contamination-subtracted visibility data, we re-image them and estimate the final root mean
square (RMS) noise levels from source-free regions. They are $18.2\rm\ \mu Jy\
beam^{-1}$ and $4.6\rm\ \mu Jy\ beam^{-1}$ in L band and  C band,
respectively.

\comment{For comparison purposes, we retrieved the Wide-field Infrared Survey Explorer (WISE) image of Pa 30 and produced the 0.3--2~keV XMM-Newton X-ray image using the observations taken in 2019 (Obs.\ ID: 	0841640101 and 084164020). The XMM-Newton data were reduced using the XMM-Newton Science Analysis System software. After filtering out the soft proton flares, the total exposure is 25.7~ks.}

\section{Results}
\label{sec:results}

\ping{The L and C bands images do not reveal significant diffuse emission from Pa~30, although a few point-like sources are found in the vicinity (see the lower panels of Figure~\ref{Fig:4in1}).}
We {marked these sources with at least $5\sigma$ detection in either of the two bands in Figure \ref{Fig:4in1}}. We used the CASA \textit{imfit} task to fit the brightness profiles of these point-like sources and found they 
can be well fitted by a 2-D Gaussian model, \commentv{with sizes and orientations consistent with the beam within $1\sigma$ uncertainties}. 
\commentv{This supports that they are indeed unresolved sources, and the corresponding} flux density measurements are summarized in Table \ref{tab:VLA_result}.

Since no significant extended structures are found within or around the field of Pa~30, we regard it as a non-detection and proceed with the analysis using a $3\sigma$ upper limit. Using the \commentv{radius} of $\sim 114.5''$, we can estimate the $3\sigma$ upper limit of Pa~30's \commentv{flux density} \ping{to be \commentv{$0.84\rm\,  mJy$} at $1.5\ \rm GHz$ and \commentv{$0.29\rm\, mJy$} at $6\ \rm GHz$. The left panel of Figure~\ref{Fig:Surfbrightness_sigmaD} compares the 1 GHz radio surface brightness $\Sigma$ and angular area between Pa~30 and known radio \commentv{shell-type} SNRs. The radio \commentv{surface brightness} of Pa~30 is $\sim 3$ orders of magnitude smaller than that of typical SNRs with comparable angular size, making it the faintest among all known Galactic SNRs.}

Moreover, there is no significant radio counterpart of the central source J005311. The RMS noise levels of the central source are $17.2\rm\, \mu Jy$ and $2.09\rm\, \mu Jy$ on L band and C band, respectively.

\section{Discussions}
\label{sec:discussions}

The absence of detectable radio emissions adds to the mystery of Pa~30, as all known young SNRs with ages less than 2 kyr in our Galaxy are radio sources \citep{Green_catlog2025JApA...46...14G}, including another Type Iax SNR at the Galactic center \citep[Sgr A East,][]{Zhou+2021ApJ...908...31Z}. Here we \ping{adopt the SNR explanation for Pa~30 and calculate the expected SNR} synchrotron emission from energy injection and particle number injection perspectives, respectively. 
These nonthermal particles can be accelerated by the SNR shocks via the diffusive shock acceleration (DSA) mechanism \citep{Bell+1978MNRAS_DSA,Blandford+1978ApJ_DSA}. 
In addition, we also provide constraints on bremsstrahlung emission and discuss the wind-blown bubble scenario.

\subsection{Synchrotron emission}

SNRs generate synchrotron radio emission as their relativistic electrons gyrate in the magnetic fields.
We assumed that the shock-accelerated particles have a power law distribution

\commentv{
\begin{equation}
    \frac{dN_i}{dE_i}=N_{0,i}\left(\frac{E_{i}}{1\,{\rm GeV}}\right)^{-s},
\end{equation}
where $i=\rm e,p$, for electron and proton, respectively. $s$ is the \commentv{particle energy spectral} index and $N_{0,i}$ is the electron or proton number.  The total energy of the particle can be determined as
\begin{equation}
    W_{i}=\int_{E_{\text{min}}}^{E_{\text{max}}}E_{i}N_{0,i}(E_{i}/1\,{\rm GeV})^{-s}dE_{i},
\end{equation}}
where $E_{\rm min}$ and $E_{\rm max}$ represent the minimum and maximum \commentv{particle} energy, respectively.
Then the synchrotron emission luminosity $L_\nu$ is mainly a function of the total electron energy $W_\text{e}$, the electron energy distribution, and the magnetic fields $B$ determined by 
\commentv{
\begin{equation}
\label{equ_LvWeB}
    L_{\nu}\propto W_{\rm e}B^{\frac{s+1}{2}}\nu^{\frac{1-s}{2}},
\end{equation}
}
In the subsequent analysis, we adopt an arbitrary maximum electron energy of $E_{\text{max}} = 10^5\,\text{GeV}$, while
\comment{varying $E_{\rm max}$ has negligible effects on the estimated luminosity when $E_{\text{max}} \gg E_{\text{min}}$.} \commentv{In this section, we adopt $E_{\text{min}}=10\,\rm MeV$  \citep{Reynolds+2021ApJ...917...55R}}. Then we can calculate the synchrotron luminosity using the synchrotron module in the python package \textit{Naima}\footnote{\url{https://github.com/zblz/naima}} \citep{Zabalza+2015Naima,Aharonian+2010PhRvD_Naima_syn}. 

The standard DSA theory predicts the \commentv{electron energy spectral} index is 2.0 for the strong shock of the SNR \citep[e.g.,][]{Bell+1978MNRAS_DSA}, which is consistent with the index distribution of the radio spectra of SNRs \citep{Green_catlog2025JApA...46...14G}. \comment{However, young SNRs show radio \commentv{frequency} spectral index ($=(s-1)/2$) from $0.6$ to $0.77$, inferring $s=2.2$--2.5. Therefore, adopting  $s=2.3$,} the 6~GHz upper limit suggests:
\commentv{
\begin{equation}
\label{equ_WeB1.5}
    \left(\frac{W_{\rm e}}{\rm erg}\right)\left(\frac{B}{10~\rm \mu G}\right)^{\frac{s+1}{2}}\lesssim 2.7\times10^{44}.
\end{equation}}
\comment{This value varies from $1.6\times10^{44}$ to $6.2\times10^{44}$ as $s$ increases from 2 to 2.5.}
Below, we calculate $L_\nu$ and also compare with \commentv{previous} models.

\commentv{If we} suppose 10\% of the \commentv{SN kinetic} energy converts to CR proton energy $W_{\rm p}$.
\ping{Adopting an electron-to-proton number ratio $K_{\rm ep}=N_{\rm e}/N_{\rm p}=0.01$ \citep[$K_{\rm ep}$ may range from $10^{-4}$ to $10^{-2}$, with $10^{-2}$ being a typical value for SNRs][]{Katz+2008JCAP} and the same particle \commentv{energy spectral index},
 we obtain $W_{\rm e}\approx 0.01W_{\rm p}=0.001E_{\rm SN}$.
This links the synchrotron \commentv{luminosity} $L_\nu$ with the \commentv{SN kinetic} energy $E_{\rm SN}$.}
\ping{On the contrary, we can derive \commentv{the upper limit of} $E_{\text{SN}}$ based on  the radio upper limit and the assumed $B$ and $E_{\rm min}$.}

For $s=2.3$, equations \eqref{equ_LvWeB} and \eqref{equ_WeB1.5} are \commentv{transformed into}  
\comment{
\begin{equation}
\label{Equ_LvESNB}
    L_{\nu}\propto E_{\text{SN}}B^{\frac{s+1}{2}}\nu^{\frac{1-s}{2}}K_{\rm ep}
\end{equation}} and the $6\,\rm GHz$ flux density upper limit provide:

\commentv{
\begin{equation}
\begin{aligned}
\label{Equ_ESNB1.5}
    E_{\text{SN}}\lesssim  2.7\times10^{47}\left(\frac{B}{10~\rm \mu G}\right)^{-1.65}\left(\frac{K_{\rm ep}}{0.01}\right)^{-1}\,{\rm erg}.
\end{aligned}
\end{equation}}

The lack of radio detection indicates \commentv{that} Pa~30 may have a low \commentv{kinetic energy} of \commentv{$E_{\text{SN}}\lesssim2.7\times 10^{47}\, \rm erg$} \citep[typically $\sim 10^{51}\rm\, erg$;][]{Vink+2020pesr.book.....V} for the typical magnetic fields $B=10$--$10^3~\mu \rm G$ in SNR shells \citep{Reynolds12}.
\ping{This is lower than ejecta kinematic energy of $\sim   10^{48}$~erg given in previous X-ray studies \citep[with large uncertainties in ambient density and ejecta mass][]{Lykou+2023ApJ...944..120L,Takatoshi2024ApJ-X-ray}}. \commentv{For $s=2$--$2.5$, and $B=3$--$10\,\rm \mu G$, the $6\,\rm GHz$ upper limit gives $E_{\text{SN}}$ between $8.5\times 10^{46}\,{\rm erg}$ and $2.6\times 10^{48}\,{\rm erg}$.}

\comment{The magnetic field strength is unknown for Pa~30, which lies 190~pc above the Galactic plane. While typical magnetic field strength in the Galactic disk is a few $\mu$G \citep{han17}, those in SNRs can be strongly amplified \citep[see a review by][]{Reynolds12}. The magnetic amplification efficiency, $\epsilon_B\equiv (B^2/8\pi)/(\rho_0 v_s^2)$, relates magnetic energy density to the postshock pressure, \commentv{where $v_s$ is the shock velocity, $\rho_0=\mu m_{\rm H} n_0$ is the ambient gas density, $\mu=1.4$ as the mean molecular weight and $n_0$ is the ambient number density.} For Pa~30, the ambient density is poorly constrained, although X-ray measurements suggest $n_0\sim 0.1~\rm cm^{-3}$ \citep{Takatoshi2024ApJ-X-ray}. This yields $\epsilon_B= 0.9\times 10^{-3} (B/10~\mu{\rm G})^2(n_0/0.1~~{\rm cm^{-3}})^{-1}(v_s/1.4\times 10^3~{\rm km\ s^{-1}})^{-2}$, falling at the lower end of $\epsilon_B=10^{-3}$--$10^{-1}$ found in young SNRs \citep{Reynolds+1981ApJ...245..912R,Reynolds+2021ApJ...917...55R}.
In the extreme case that Pa~30 lies in a very rarefied region and has an ultra-low magnetic field $B\sim 3\mu$G, the radio measurements still suggest a low \commentv{kinetic energy} $\lesssim 10^{48}$~erg.
}

The low \commentv{SN kinetic} energy can explain the peculiar position of Pa~30 at the lower-left corner in the $\Sigma$--D diagram (right panel of Figure~\ref{Fig:Surfbrightness_sigmaD}). 
Previous studies have shown a positive correlation between the radio \commentv{surface brightness} $\Sigma$ and  $E_{\rm SN}$  \citep{Berezhko+2004A&A..Sigma-D, Pavlovic+2018ApJ.Sigma-D}.
\comment{One can predict $\Sigma$--D relation of SNRs by combining Equation~\ref{Equ_LvESNB} and the SNR dynamic evolution equations.
We use the self-similar solutions of \cite{Truelove_1999ApJS..120..299T} for describing the dynamic evolution of SNRs in the ejecta-dominated phase and Sedov-Taylor phase. Pa~30 is probably in the ejecta-dominated phase with little velocity deceleration \citep{Cunningham2024ApJL_age}. Therefore, we adopt the model with a uniform ejecta distribution, while the models with a sharp ejecta density profile (power-law index $n > 3$) predict significant deceleration in the ejecta-dominated phase and are thus inconsistent with the case in Pa~30. }

In the ejecta-dominated phase, the shock velocity scales as 
\begin{equation}
    v_s \propto E_{\rm SN}^{1/2} M_{\rm ej}^{-1/2},
\end{equation}
Combining this with equation~\ref{Equ_LvESNB}, \comment{adopting $s=2.3$}, we derive
\begin{equation}
\Sigma=L_\nu(\pi^2 D^2)^{-1} \propto E_{\text{SN}}^{3.3/4} M_{\rm ej}^{0.7/4} n_0^{3.3/4} \epsilon_B^{3.3/4}K_{\text{ep}}.
\end{equation}

The right panel of Figure~\ref{Fig:Surfbrightness_sigmaD} shows the $\Sigma$-$D$   relationship of some \commentv{shell-type} SNRs and our prediction. The observed values of SNRs can be reproduced by our model with normal SNR parameters: \commentv{$s=2$--$2.5$}, $E_{\text{SN}}=10^{50}$--$2\times10^{51}\,\rm erg$, ejecta mass $M_{\rm ej}=1$--$4M_{\odot}$, $n_0=0.05$--$1\,\rm cm^{-3}$, and $\epsilon_{B}=0.001$--\commentv{$0.1$}.  Pa~30 is obviously an outlier in this figure.
Given the $3\sigma$ radio upper limit of Pa~30 and $v_s\simeq1400\,\rm km\,s^{-1}$, our model yields the constraint
\commentv{
\begin{equation}
    \left(\frac{E_{\text{SN}}}{10^{48}\,\rm erg}\right)\left(\frac{n_0}{0.1\,\rm cm^{-3}}\right)^{3.3/4} \left(\frac{\epsilon_B}{0.001}\right)^{3.3/4}\left(\frac{K_{\text{ep}}}{0.01}\right) \lesssim 0.2.
\end{equation}
}

For $\epsilon_{B}=0.001$, the location of Pa~30 in the $\Sigma$--$D$ diagram suggests a \commentv{kinetic energy $\lesssim 2\times10^{47}\, \rm erg$ if $n_0=0.1~\rm cm^{-3}$, or $\lesssim 2\times10^{48}\,\rm erg$ if $n_0=0.006~\rm cm^{-3}$}, which is over 3 orders of magnitude lower than the canonical \commentv{SN kinetic} energy. We note that the above calculation is based on the assumption that $\sim 10\%$ \commentv{SN kinetic} energy is converted to CRs, which is a conventional number for SNRs \citep{Vink+2020pesr.book.....V}. However, it may not be valid for the newly formed SNRs since CR acceleration needs time, and this number is to be tested for Pa~30.

\commentv{\subsection{Comparison with previous model and recalculation of $\epsilon_e$}}
\label{subsubsec:Compare with previous}

Considering the atypical morphology of Pa~30 compared to normal SNRs, \cite{Takatoshi2024ApJ-X-ray} constructed a theoretical model including the inner shock power by the termination shock from the WD wind and the outer shock from the SNR. They estimated the synchrotron emission spectra under the hypotheses of different electron acceleration efficiency $\epsilon_\text{e}$. They found that the sensitivities of the archive data (i.e. $4.13\, \rm mJy$ at $1.4\,\rm GHz$ for NVSS, $20.4\, \rm mJy$ at 2--4$\,\rm GHz$ for VLASS, and $9\, \rm mJy$ at $408\,\rm MHz$ for CGPS) cannot tightly constrain parameters. 
With deep, targeted observation in L and C bands, we refined the $3\sigma$ upper limits to \commentv{$0.84\rm\,  mJy$} in L band and \commentv{$0.29\rm\, mJy$} in C band.

With our new data, we can rule out the previously assumed optimistic luminosity of the outer shock (see Figure \ref{Fig:Synchro_spec}
) and suggest a smaller electron acceleration efficiency $\epsilon_\text{e}$ or smaller postshock gas number density $n_0$ of the shock region. 
\comment{The equations used for the calculation of $\epsilon_e$ is elaborated in \cite{Takatoshi+2024PASJ_radio}.} For $\epsilon_B=0.01$ used by \cite{Takatoshi+2024PASJ_radio}, our observations suggests $\epsilon_\text{e}(\frac{n_0}{0.1\,\rm cm^{-3}})^{\frac{s+5}{4}}\lesssim4.0\times10^{-4},3.0\times10^{-3}, 7.9\times10^{-2}$ for $s=2,2.5$, and 3, respectively.
\comment{Adopting $\epsilon_B=0.001$ , we obtained $\epsilon_e(\frac{n_0}{0.1~\rm cm^{-3}})^{\frac{s+5}{4}}\lesssim 2.5\times10^{-3},2.2\times10^{-2},0.78$ for $s=2,2.5$, and 3, respectively.} \commentv{This model assumes a uniform distribution of CR electrons, while a shell-like geometry could change the results by an order of magnitude. A detailed discussion is out of the scope of this letter.}

\subsection{Bremsstrahlung}
\commentv{Our radio observations have not established the existence of synchrotron emission from the SNR shock. }
\commentv{Previous optical study suggested that Pa 30's unusual appearance with radially aligned filaments may arise from the photoionization of wind-driven ejecta \citep{Fesen+2023ApJ...945L...4F}. In this case, the hot, ionized bubble could produce thermal bremsstrahlung emission.}
The non-detection of radio emission from Pa~30 thus sets an upper limit on this process and offers constraints on ionized gas. 
\commentv{Assuming that the thermally emitting gas is dispersed within a sphere of radius $R$, and the volume filling factor is $f$.}
\commentv{Then, the bremsstrahlung intensity of an optically thin, spherically distributed gas is calculated as 
$I_\nu\approx\frac{2}{3}B_\nu \tau$}
where $\tau$ is the optical depth measured across the center 
of the nebula
\citep{Osterbrock+1965ApJ...141.1285O,Olnon+1975A&A....39..217O_Bremss_absorp}. 
Considering Rayleigh–Jeans regime ($h\nu \ll kT$)
, we obtain \commentv{the bremsstrahlung luminosity}
\commentvv{
\begin{equation}
    \begin{aligned}
    L_\nu&\approx 1.9\times 10^{18}\sum\limits_{\rm i} Z_{\text{i}}^2n_{\text{i}}n_{\rm e}f\left(\frac{T}{\text{K}}\right)^{-0.35}\left(\frac{\nu}{\text{GHz}}\right)^{-0.1} \\&{\rm\,erg\,s^{-1}\,Hz^{-1}},
\end{aligned}
\end{equation}
}
where $Z_{\text{i}}$ \commentvv{and} $n_{\text{i}}$ represent the charge and number density of the $i$-th \commentvv{ion specie}, respectively.

For the fully ionized, solar abundance gas (\commentvv{dominated by H and He}) that fully fills the nebula, 
the radio observation sets a maximum density of \commentv{$n_\text{e}=1.6\sim5.3~\rm cm^{-3}$} for $T=10^4\sim10^7$~K. 
It is more likely that Pa~30 consists of weakly ionized gas that fills only a fraction of the volume and has over-solar abundances, since the optical structures are filamentary and weak in H$\alpha$ emission 
\citep{Fesen+2023ApJ...945L...4F}. 
Pa~30 is also filled with hot, highly ionized plasma, which emits the bremsstrahlung emission mainly in the X-ray bands, rather than in the radio bands.
Therefore, we consider the radio-emitting gas to be weakly ionized ($Z_{\text{i}}=1$ \commentvv{and $n_{\rm e}=\sum\limits_{\rm i}n_{\rm i}$}), as also inferred by the strong [S II], [N II], and [O II] lines.
Given the 0.1~mJy upper limit in the C band, 
we estimate \commentv{
\begin{equation}
    f(\frac{n_{\rm e}}{\rm cm^{-3}})^2(\frac{T}{\rm K})^{-0.35}\lesssim0.1.
\end{equation}}
Using the average density derived from [S II] lines  $n_{\rm e}=120\, \rm cm^{-3}$ \citep{Lykou+2023ApJ...944..120L}, the filling factor of the weakly ionized gas is derived as \commentv{$f\lesssim 2\times10^{-4} (T/10^4~\rm K)^{0.35}$.} 
Therefore, the lack of radio emission suggests that the optical structures may fill less than \commentv{$\sim 2\times10^{-4}$} \commentv{of the total nebula's volume. This is consistent with the unresolved, filamentary structures in the optical band \citep{Fesen+2023ApJ...945L...4F}. }
\comment{However, the actual temperature is uncertain and may deviate from the $10^4\,\rm K$, introducing extra uncertainty in the estimate of the filling factor. }

\commentv{We can roughly estimate the upper limit of the total mass of the gas that could produce radio bremsstrahlung radiation as} 
\begin{equation}
    M_{\rm gas}\lesssim5\times 10^{-2}M_{\odot}\left(\frac{T}{10^4\,\rm K}\right)^{0.35}\left(\frac{n_{\rm e}}{120\,\rm cm^{-3}}\right)\left(\frac{\mu}{10}\right),
\end{equation}
\commentv{where a large mean molecular weight is assumed to count for the metal-rich gas. Considering the metal composition and the ambient density carry large uncertainties, we therefore do not discuss the gas mass further.}

\subsection{Sub-energetic SN origin of Pa 30 or not?}

\ping{Pa~30 has been regarded as a young SNR, although this interpretation remains unconfirmed. 
This nebula challenges our understanding of SNe/SNRs in several aspects if it is indeed an SNR.
Unlike a normal Ia SNR without a central source or core-collapse SNR with a neutron star or black hole, it consists of a central source exhibiting ultra-strong outflow (16,000 km~s$^{-1}$) and the mass-loss rate ($\dot{M}$) of $3.5\times10^{-6}M_{\odot}\,\rm yr^{-1}$ \citep{Gvaramadze+2019Natur.569..684G}.
A Type Iax origin has therefore been suggested, as this subclass of subluminous SNe may leave behind a hot, partially unburnt white dwarf \citep{Oskinova+2020A&A}.

Additionally, the absence of detectable radio emission renders Pa~30 the least luminous known SNR in the radio band.  It is particularly puzzling that an 844-year-old remnant is so faint at radio wavelengths, while other young SNRs are typically bright radio sources  (see Figure~\ref{Fig:Surfbrightness_sigmaD}). 
The extremely low radio synchrotron luminosity implies that few CR electrons are being accelerated and suggests an unusually low \commentv{kinetic energy of $\lesssim 2.7\times 10^{47}(B/10~\mu\rm G)^{-1.65}$~erg} ---  3--4 orders of magnitude lower than the canonical \commentv{SN kinetic} energy.

Since most Galactic SNRs have been identified through radio surveys, radio-faint remnants like Pa~30 are likely to be missed unless detected in other wavebands. This introduces an observational bias, implying that the current Galactic SNR sample is highly incomplete and that a larger population of sub-energetic SNRs remains undiscovered. \commentv{Such low-kinetic-energy SNRs are intrinsically difficult to detect, but they may constitute a non-negligible fraction of the Galactic SNR population. Since Pa 30 may link to a historical supernova, this implies that such events are not extremely rare. In fact, type Iax supernovae are estimated to account for $\sim 10\%$ of Type Ia explosions \citep[][see \cite{Foley+2013ApJ...767...57F} for a larger fraction]{MA+2025A&A...698A.305M}. Moreover, given the non-detection at such a close distance, similar remnants located farther away in the Galaxy would likely remain undetected \citep{Ranasinghe+2022ApJSNRDistance}.}  Identifying these remnants is crucial for understanding the true distribution of \commentv{SN kinetic} energy and metal feedback into the Galaxy, as well as the diverse outcomes of stellar or white dwarf evolution. 

On the other hand, the low \commentv{kinetic energy} further challenges our definition of SNe, which are canonically characterized by a \commentv{kinetic energy} of $\sim 10^{51}$~erg, even though a population of sub-energetic SN is emerging \citep[e.g., Type Iax SN 2008ha with $E_{\rm SN}\sim 2\times 10^{48}$~erg,][]{Foley09}.

Therefore, it remains unclear whether Pa~30 is a remnant of a Type Iax SN or a wind nebula powered by the central source. For the latter case, our calculation of the synchrotron emission is inappropriate since it is based on SNR parameters.
Notably, the central source of Pa~30 is an energetic source and should at least partially shape Pa~30. It gives a wind luminosity of $L_w=1/2 \dot{M} v_w^2=2.8\times10^{38}\,\rm erg\,s^{-1}$ and can supply an energy of $10^{48}$~erg in $\sim110$ years. However, it is known that the wind luminosity is not constant over time \citep{Schaefer+2023MNRAS}.
There is also a possibility that Pa~30 is a complex nebula powered by both an earlier explosion and the central source.
We anticipate that future multi-wavelength observations will unveil the radiative mechanism and material composition, thereby shedding light on the nature of Pa~30.}

\section{Summary}
\label{sec:summary}
We report the VLA observation of Pa~30,
\ping{which hosts a central source with extraordinarily fast outflow.}
Our observations \commentv{in} L band and C band resulted in the non-detection of shell-like extended structures that resemble their infrared or X-ray morphology, at the rms noise level of $18.2\, \rm \mu Jy\, beam^{-1}$ in L band and $4.6\, \rm \mu Jy\, beam^{-1}$ in C band. Some of the bright radio sources in the Pa~30 field are likely irrelevant background radio-emitting objects. We employed different methods to constrain the properties of Pa~30 using the upper limits on the radio \commentv{flux density}. 

First, we considered the SNR scenario of Pa~30 and 
our radio observation provides an independent way to infer the low \commentv{kinetic energy to $E_{\text{SN}}\lesssim 2.7\times10^{47}(B/10\,\rm \mu G)^{-1.65}\, \rm erg$.}

We also considered the wind nebula \ping{case and the radio emission is } dominated by bremsstrahlung emission of thermal electrons. Our radio observation helped to provide constraints on the gas density and filling factor.

While our radio observations support Pa~30's being a remnant of a peculiar, sub-energetic supernova explosion, other scenarios, such as a hot stellar bubble, remain plausible. Deep radio exploration and multi-wavelength observations of Pa 30 and other potential sub-energetic SNRs could provide definitive characterization.

\facilities{VLA, XMM-Newton}

\software{astropy \citep{2013A&A...558A..33A,2018AJ....156..123A,2022ApJ...935..167A},  
          CASA \citep{CASA2022PASP..134k4501C},
          CARTA \citep{Comrie+2021_CARTA},
          SAS \citep{GabrielSAS2004ASPC..314..759G}
          }

\begin{acknowledgments}
We thank the anonymous reviewer for the constructive comments. This work was supported by the National Natural Science Foundation of China 
Nos.\ 12273010, 12588202, 12041301, 12173018, 12121003, and 12393852. P.Z.\ acknowledges the support from the China Manned Space Program with grant No.\ CMS-CSST-2025-A14.
\end{acknowledgments}

\appendix
\section{Synchrotron radiation model by T.Ko et al. (2024A)}
\begin{figure*}[htp]
\centering
\includegraphics[width=0.95\textwidth]{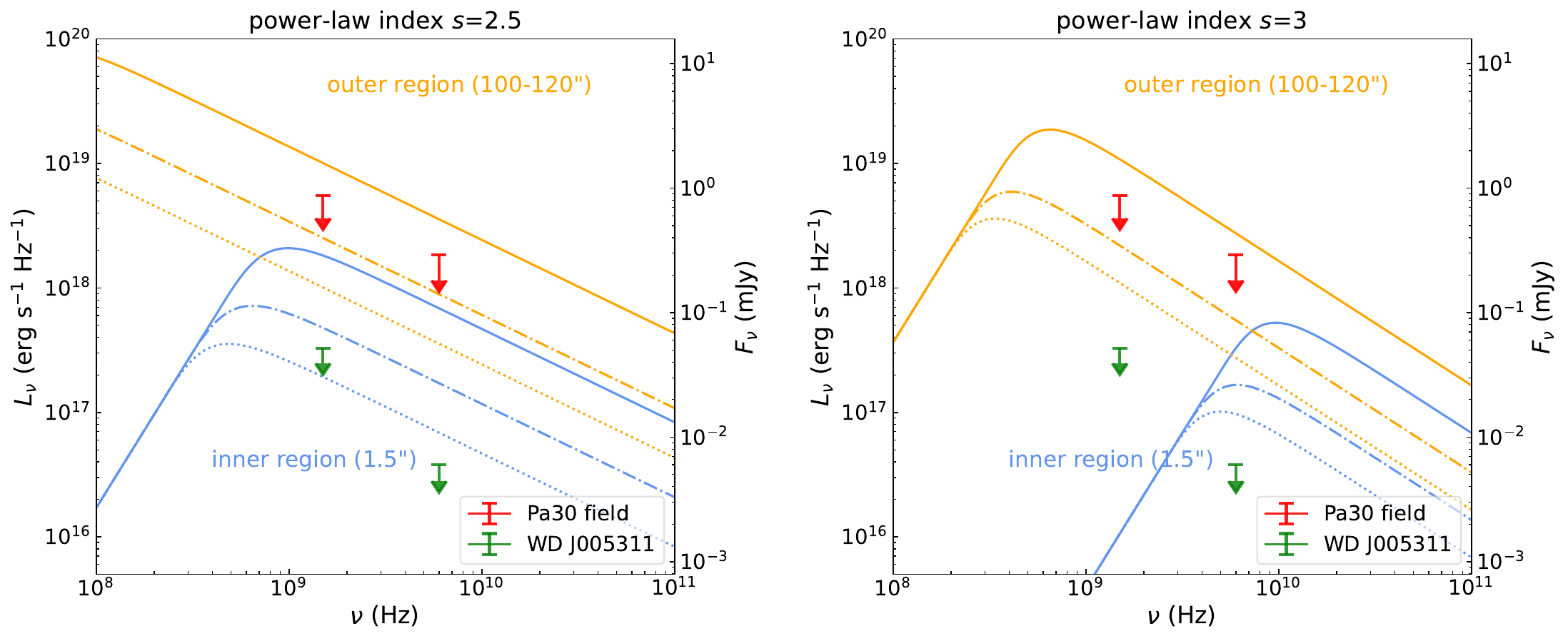}
\caption{Synchrotron spectra predicted by the model of \citep{Takatoshi+2024PASJ_radio} with electron \commentv{energy spectral index} $s=2.5$ and $s=3$, respectively. The orange and blue lines represent the outer region (SNR shock) and inner region (termination shock), respectively. The solid, dash-dotted, and dotted refer to different electron acceleration efficiencies (corresponding to optimistic, fiducial, and pessimistic assumptions, respectively): $\epsilon_\text{e}=2\times10^{-3},\ 5\times10^{-4},\ \text{and } 2\times10^{-4}$ for $s=2.5$; $\epsilon_\text{e}=5\times10^{-2},\ 1\times10^{-2},\  \text{and } 5\times10^{-3}$ for $s=3$, respectively. The red and green arrows show the upper limit of the Pa~30 field and the central source J005311 in this work.}
\label{Fig:Synchro_spec}
\end{figure*}

\commentv{Figure \ref{Fig:Synchro_spec} compares the radio upper limits of Pa~30 and the central source  with the models by \cite{Takatoshi+2024PASJ_radio}. In their model, the minimum and maximum electron energies are assumed to be $E_{\text{min}}=2m_{\rm e}c^2$ and $E_{\text{max}}=10^7m_{\rm e}c^2$, respectively, and $\epsilon_{\rm B}=0.01$ is adopted. Only $\epsilon_\text{e}(\frac{n_0}{0.1\,\rm cm^{-3}})^{\frac{s+5}{4}}\lesssim3.0\times10^{-3}$ for $s=2.5$, and $\lesssim7.9\times10^{-2}$ for $s=3$ can satisfy our observation.}

\bibliography{sample7}{}
\bibliographystyle{aasjournalv7}

\end{document}